\documentclass[aps,prb,twocolumn,showpacs,superscriptaddress,groupedaddress]{revtex4}

\usepackage{graphicx}
\usepackage{dcolumn}
\usepackage{subfigure}
\usepackage{wrapfig}
\usepackage{cancel}
\usepackage{color}
\usepackage{textcomp}
\usepackage{amsmath}
\usepackage{amsfonts}
\usepackage{amssymb}
\usepackage{array}
\usepackage[utf8]{inputenc}

\newcommand{\linio}{LiNiO$_2$}
\newcommand{\nanio}{NaNiO$_2$}
\newcommand{\jt}{Jahn-Teller}

\newcommand{\etal}{\it{et~al.}}
\newcommand{\abinitio}{\it{ab initio}}

\DeclareUnicodeCharacter{2212}{-}

\begin{document}

\title{LiNiO$_2$ as a high-entropy charge- and 
bond-disproportionated glass}

\author{Kateryna Foyevtsova$^{\ast}$}
\affiliation{Department of Physics \& Astronomy, University of British Columbia,
Vancouver, BC V6T 1Z1, Canada}
\affiliation{Stewart Blusson Quantum Matter Institute, University of British Columbia,
Vancouver, BC V6T 1Z4, Canada}
\author{Ilya Elfimov}
\affiliation{Department of Physics \& Astronomy, University of British Columbia,
Vancouver, BC V6T 1Z1, Canada}
\affiliation{Stewart Blusson Quantum Matter Institute, University of British Columbia,
Vancouver, BC V6T 1Z4, Canada}
\author{Joerg Rottler}
\affiliation{Department of Physics \& Astronomy, University of British Columbia,
Vancouver, BC V6T 1Z1, Canada}
\affiliation{Stewart Blusson Quantum Matter Institute, University of British Columbia,
Vancouver, BC V6T 1Z4, Canada}
\author{George A. Sawatzky}
\affiliation{Department of Physics \& Astronomy, University of British Columbia,
Vancouver, BC V6T 1Z1, Canada}
\affiliation{Stewart Blusson Quantum Matter Institute, University of British Columbia,
Vancouver, BC V6T 1Z4, Canada}

\date{\today}

\begin{abstract}
Understanding microscopic properties
of {\linio},
a Li-ion battery cathode material with
extraordinarily high reversible capacity,
has remained a challenge for decades.
Based on extensive electronic structure calculations,
which reveal a large number of nearly degenerate
phases involving local Jahn-Teller effect as well as
bond and oxygen-based charge disproportionation,
we propose that {\linio} exists in a high-entropy
charge-glass like state at and below
ambient temperatures.
Recognizing the glassy nature of {\linio}
does not only explain its key experimental features,
but also opens a new path in designing entropy-stabilized
battery cathodes with superb capacities.
\end{abstract}

\maketitle 

\section{Introduction}

LiNiO$_2$ is well known as a promising cathode
material for rechargeable Li-ion
batteries\cite{Goodenough55,Goodenough58}.
One of the most fascinating but also puzzling properties of this oxide is its
record-breaking stability upon repeated charge and discharge cycles.
It is indeed quite surprising that the {\linio} system can be cycled more than a few
times changing the Li concentration by up to 80\%
on each cycle\cite{Liu15,Arai95}.
This indicates that there must be something very special about the
crystal structure of {\linio}
that keeps the original structure in place, and indeed
{\linio} has demonstrated a range of structural
features that are hard to understand in a conventional way.

\begin{figure*}
\begin{center}
\includegraphics[width=0.98\textwidth]{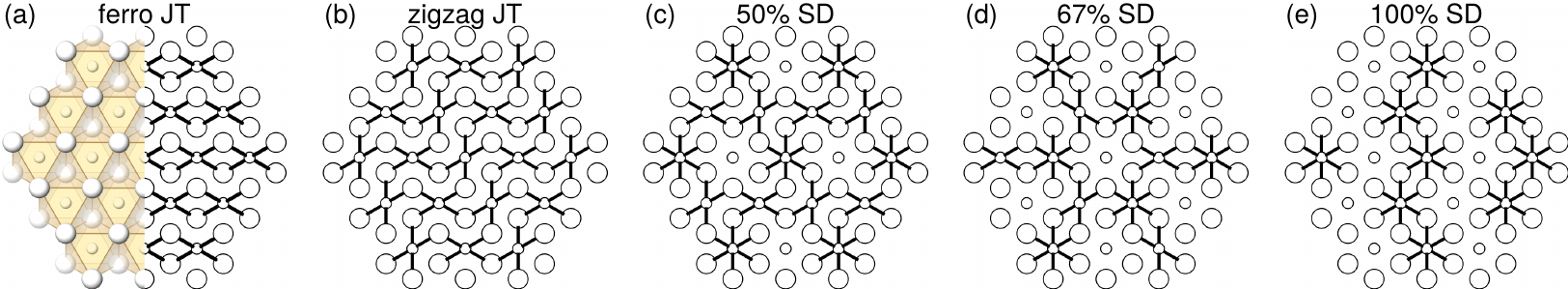}
\caption{Possible orderings of short
and long Ni-O bonds in the triangular lattice of NiO$_6$ octaherda.
Nickel (oxygen) atoms are represented by small (large) circles.
The short Ni-O bonds are shown as black connecting lines.
In (a), a three-dimensional view of the lattice
is shown on the left-hand side for a reference.
Panels (a) and (b) represent phases where only
Jahn-Teller (JT) distorted NiO$_6$ octahedra are present,
whereas panels (c), (d), and (e) represent phases
where, respectively, 50\%, 67\%, and 100\% of the
NiO$_6$ octahedra are size-disproportionated (SD).}
\label{Fig:str}
\end{center}
\end{figure*}

{\linio} consists of layers of
edge-sharing NiO$_6$ octahedra on a triangular lattice
alternating with Li layers.
The formally trivalent Ni ions are in a low-spin
configuration $t_{2g}^6e_g^1$ which makes them
{\jt} (JT) active\cite{Goodenough58,Hirakawa85,Hirota91,Yamaura96}.
Despite this, however, the NiO$_6$ octahedra
do not undergo a {\it cooperative} JT distortion even at
the lowest temperatures,
as follows from the numerous Rietveld refinements of
the {\linio} powder diffraction
data\cite{Goodenough58,Arai95,Chung05}.
This is in a surprising contrast with the behavior of its sister
compound {\nanio} that
shows a ferro-orbital ordering of occupied $d_{z^2}$
orbitals and an associated cooperative elongation of the NiO$_6$
octahedra below 460~K
as illustrated in Fig.~\ref{Fig:str}(a)\cite{Bongers66,Holzapfel04}.
On the other hand, structural
analyses based on the neutron pair distribution function (nPDF)\cite{Chung05},
extended x-ray absorption fine structure\cite{Rougier95},
and electron spin resonance\cite{Barra99,Reynaud01}
measurements, strongly suggest
{\it local}, but apparently {\it disordered,} Ni-O bond disproportionation
in {\linio}.
Disorder effects in this system,
which also include a spin-glass-like behavior
at low temperatures\cite{Hirota91},
are often linked to its unavoidable deviation from
stoichiometry\cite{Arai95,Petit06}.
Yet, the exact nature of the disordered state in {\linio}
has so far remained unclear.

Apart from the above,
{\linio} also possesses a rather peculiar electronic property,
which, although being often dismissed in literature,
might be the key to understanding its structural properties.
As was first pointed out by Kuiper {\it et al}.\cite{Kuiper89},
this system is in the negative charge-transfer regime
which brings the NiO$_6$ octahedra closer
to the $(t^6_{2g}e^2_g)\underline{L}$ (rather than the formal $t^6_{2g}e^1_g$)
{\it average} configuration,
where $\underline{L}$ denotes a ligand hole residing
on a molecular orbital formed by the six oxygen-$2p_{\sigma}$
orbitals in an octahedron.
In this regime, the {\jt} effect
might be in competition with another degeneracy lifting mechanism,
observed in rare-earth nickel perovskites\cite{Johnston14,Green16,Foyevtsova15},
which involves formation of two types of size-disproportionated
NiO$_6$ octahedra with configurations
$t^6_{2g}e^2_g$
and $(t^6_{2g}e^2_g)\underline{L}^2$ with
symmetries
and spins corresponding to those of Ni$^{2+}$ and Ni$^{4+}$.
Not surprisingly, therefore, a strongly competing bond-disproportionated phase of {\linio},
with ordered stripes
of collapsed and expanded
NiO$_6$ octahedra [Fig.~\ref{Fig:str}(e)],
was indeed recently predicted within density functional theory (DFT)\cite{Chen11}.

Our present theoretical study
offers a unifying explanation of the puzzling structural and electronic
properties of {\linio} outlined above
as well as its superior performance in Li-ion batteries
by recognizing
the fundamental role of entropy in this system.
Based on extensive electronic structure calculations, we
demonstrate that {\linio}
has a large number of nearly degenerate states
involving NiO$_6$ octahedra with $t^6_{2g}e^2_g$, $(t^6_{2g}e^2_g)\underline{L}$
and $(t^6_{2g}e^2_g)\underline{L}^2$ configurations,
leading to basically a large
number of combinations of these into glassy-like structures which increases
the entropy resulting in a high-entropy material.
It is shown that, as a result of the
charge-transfer energy being negative in {\linio},
this state involves charge disproportionation
on oxygen rather than on nickel atoms.
Although entropy is known to play important roles in polymer
science and protein folding as well as glasses, multi-component oxides,
and metal alloys\cite{Rost15,Yao18}, these
are mostly structurally driven while our proposal involves an
electronic-structure driven mechanism for the pure material.

\section{Methods}
We performed density functional theory (DFT) calculations with the
pseudo-potential code
VASP\cite{Kresse96,Paier05} and the all-electron code
WIEN2k\cite{wien2k}.
All of our structural relaxations were done in VASP,
using a high energy cut-off of 550~Ry
and fine $k$-grids with densities close to that of
a $12\times12\times12$ $k$-grid in a
$5.5\AA\times5.5\AA\times5.5\AA$ unit cell.
A combination of the local density approximation (LDA)\cite{Perdew92}
and the LDA+U method as designed by
Anisimov~{\it et al.} \cite{Anisimov91,Anisimov93,Liechtenstein95},
with the Ni-$3d$ electrons' on-site
interactions $U=6$~eV
and $J_{\text{H}}=1$~eV,
was used to treat  exchange and correlation effects.

The neutron pair distribution functions (nPDF), $G(r)$,
of the relaxed {\linio} phases were simulated with the
PDFfit2 program\cite{pdfgui} using the following expression\cite{Chung05}:
\[
G(r) = \frac{1}{rN}\sum_{i,j}
\left[
\frac{b_ib_j}{\langle b \rangle^2}
\delta(r-r_{ij})
\right]
- 4\pi \rho_0,
\]
where $r$ is distance, $b_i$ and $r_{ij}$
are the scattering length of the $i$th atom and the distance
between the $i$th and $j$th atom, respectively, $N$ is the
number of atoms, and $\rho_0$ is the average atomic density.
Thermal broadening was accounted for through
introducing the Debye-Waller factors.
We used theoretical structural parameters, scaled with
an empirical factor of 1.024, and theoretical
thermal displacement parameters. In the PDFfit2 program,
we chose 0.08 for the $Q_{\text{damp}}$ parameter and 0.5 
for the linear atomic correlation factor $\delta_1$ for
all the considered model {\linio} phases except the
rhombohedral one where we used $\delta_1=0.2$.
The neutron time-of-flight (TOF) powder diffraction patterns
were simulated using the GSAS package\cite{gsas,expgui}.
For both the nPDF and the TOF powder diffraction simulations,
we used the following theoretically calculated atomic
isotropic displacement parameters $U\equiv U_{11}=U_{22}=U_{33}$,
obtained for the temperature of 10~K employing VASP and the Phonopy
package\cite{phonopy}: $U$(Li) = 0.008~$\AA^2$, 
$U$(Ni) = 0.0011~$\AA^2$,
$U$(O) = 0.0025~$\AA^2$.
The thermodynamic properties
were calculated using the phonon spectra obtained with VASP and Phonopy via
the dynamical matrix method.

\section{Results}
Our following discussion will be based on a comparison of
a number of different bond-disproportionated phases of {\linio}
shown in Fig.~\ref{Fig:str}, which all turn out to
be easily achievable in a DFT structural relaxation.
They include two JT phases [(a) and (b)],
the stripy phase with size-disproportionated NiO$_6$ octahedra
from Ref.~\onlinecite{Chen11} (e), and two mixed phases where
JT-distorted and size-disproportionated octahedra
are simultaneously present [(c) and (d)].
The high-symmetry
rhombohedral $R\bar{3}m$ phase of {\linio},
which has been traditionally assigned to {\linio}
based on powder diffraction refinement and has no Ni-O bond disproportionation,
will also be considered for completeness.
For each phase, we use DFT and the local density approximation $+U$ (LDA+U) method
to perform full lattice
relaxation and to calculate the ground state electronic structure.
Ferromagnetic alignment of Ni spins is
adopted throughout all the calculations, as
it was established that
the total energy differences between different
spin configurations are much smaller than those
between the different structural phases of {\linio},
as detailed in the Appendix.
This finding of ours agrees with the conclusions of
Mostovoy and Khomskii\cite{Mostovoy02} who demonstrated decoupling
of spin and orbital degrees of freedom in {\linio}.

\begin{table}
\begin{center}
\begin{tabular}{lcccc}
\hline\hline
Phase & $E_{\text{tot}}$ (meV/f.~u.)  & Gap (eV)  &
$N_{\text{short}} / N_{\text{long}}$&
$d_{\text{Ni-O}}$ (\AA) \\
\hline
Rhomb.            & 141.3 (147.5)  & 0    & -   &   1.98    \\
ferro JT          &  \hspace{-0.53cm}18.1 (0)      & 0.43 & 2   & 1.90 \, 2.12 \\
zigzag JT         &  \hspace{0.53cm}0   (19.1)   & 0.56 & 2   & 1.90 \, 2.11 \\
50\% SD           &  49.6 (60.1)   & 0.35 & 1.4 & 1.90 \, 2.06 \\
67\% SD           &  32.1 (43.5)   & 0.51 & 1.25& 1.90 \, 2.05 \\
100\% SD          &  26.3 (63.3)   & 0.58 & 1   & 1.89 \, 2.06 \\
\hline\hline
\end{tabular}
\end{center}
\caption{Calculated properties
of the {\linio} model phases.
$E_{\text{tot}}$ are relative total energies,
$N_{\text{short}}/N_{\text{long}}$ are
the ratios of the numbers of short and long
Ni-O bonds,
and $d_{\text{Ni-O}}$ are the average nearest-neighbor
Ni-O bond lengths.
A scaling factor of 1.024 has been applied to the
original $d_{\text{Ni-O}}$
values obtained within LDA+U. For comparison, $E_{\text{tot}}$ values relevant
for {\nanio} are provided in parentheses.
}
\label{T:comp}
\end{table}

\begin{figure}
\begin{center}
\includegraphics[width=0.47\textwidth]{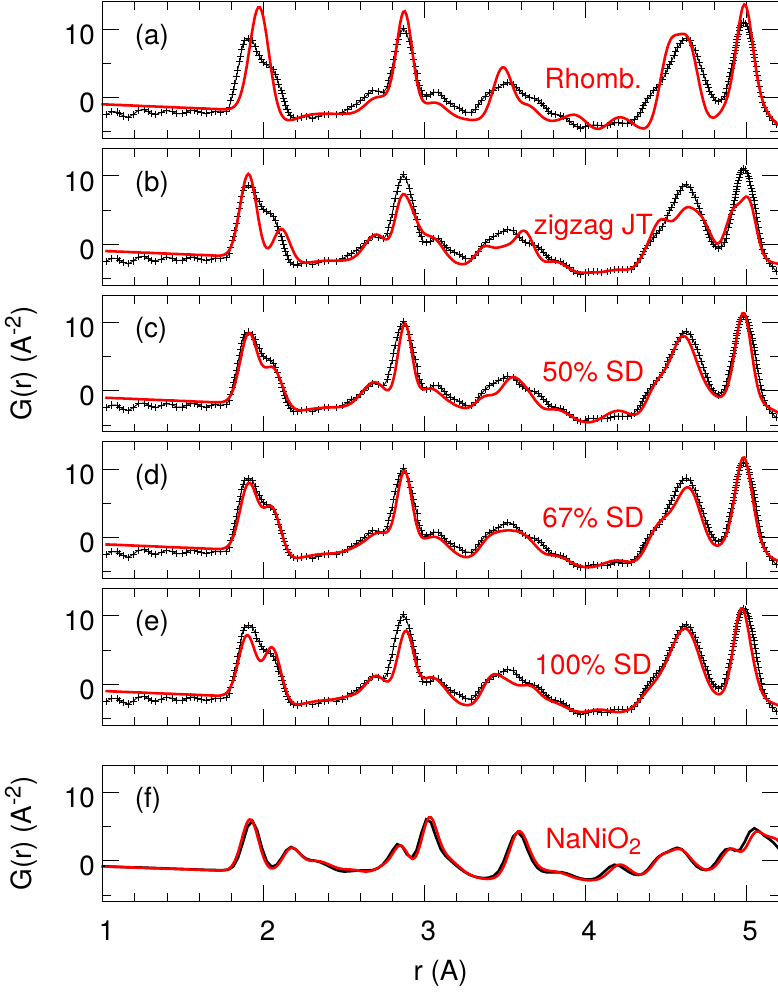}
\caption{
Neutron pair distribution functions (nPDF) $G(r)$ of {\linio}
and {\nanio}. In (a)-(e),
the experimental (black symbols)\cite{Chung05} and various
theoretically simulated (red lines)
nPDFs at the temperature of 10~K are compared for {\linio}.
Panel (f) shows nPDFs of {\nanio} simulated using
its experimental crystal structure (black line)\cite{Holzapfel04}
and using a relaxed structure from LDA+U calculations (red line).
A scaling factor of 1.024 has been applied to $r$
in $G(r)$ calculated using results of LDA+U structural relaxations.}
\label{Fig:PDF}
\end{center}
\end{figure}

\begin{figure}
\begin{center}
\includegraphics[width=0.47\textwidth]{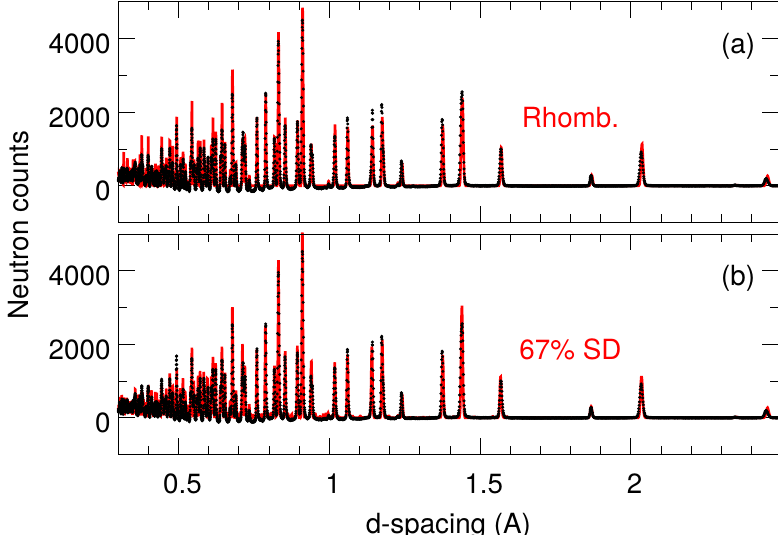}
\caption{Neutron time-of-flight
powder diffractograms of {\linio}.
Experiment  (black symbols)\cite{Chung05} is compared to 
a simulation (red lines) where (a) the rhombohedral
and (b) the 67\% SD model phase of {\linio} at 10~K
is used. A
scaling factor of 1.024 has been applied to the $d$-spacing
in the simulated diffractograms.}
\label{Fig:powder}
\end{center}
\end{figure}

We note first that for {\nanio} (the sister compound)
DFT correctly predicts the JT phase
with a ferro-orbital ordering of occupied $d_{z^2}$ orbitals
[Fig.~\ref{Fig:str}(a)]
to be the lowest energy state\cite{Meskine05,Chen11}.
For {\linio}, on the other hand, we find
that the lowest energy state
is the JT phase with a zigzag ordering of occupied $d_{z^2}$ orbitals
[Fig.~\ref{Fig:str}(b)]\cite{Chen11}. Its energy is by 18.1~meV per formula
unit (f.~u.) lower than that of the ferro-orbitally ordered phase
and by 141.3~meV/f.~u. lower than that of the rhombohedral phase
(see Table~\ref{T:comp}).
The fully and partially size-disproportionated (SD) phases shown in
Figs.~\ref{Fig:str}(c)-(e)
have energies
that are not too far from the energy of the ferro-orbital
JT phase.
For example, the three-fold rotationally symmetric
phase shown in (d), with two thirds (or 67\%) of Ni sites being SD, is higher
in energy by only 14~meV/f.~u.
As one can also see in Table~\ref{T:comp},
all the considered bond-disproportionated phases
have a small charge gap in LDA+U,
which qualitatively agrees with experiment\cite{Hirota91}.
Interestingly, the same SD phases
can be obtained for
{\nanio}, but, as also shown in Table~\ref{T:comp},
they are energetically more strongly removed from the
lowest energy state than in the case of {\linio}.

Before discussing the important implications
following from these total energy calculations, let us point out
that {\linio} being in a mixed phase, such as the 67\% SD one,
could explain amazingly well the findings from both,
the pair distribution function and the powder diffraction analyses,
which previously have been regarded as mutually contradicting.
In Figs.~\ref{Fig:PDF}(a)-(e), the experimental
nPDF of {\linio}\cite{Chung05}
is compared with simulated
nPDFs of the five model phases from Fig.~\ref{Fig:str}.
Both the experiment and the simulations
are performed at the temperature of 10~K.
In order to correct for the underestimation of
interatomic distances in LDA+U,
an empirical scaling factor of 1.024
was applied to $r$ in the simulated nPDF $G(r)$.
This procedure
gives very good agreement between theory and experiment for {\nanio},
whose experimental crystal structure is unambiguously
known [Fig.~\ref{Fig:PDF}(f)]\cite{Holzapfel04}.
For {\linio}, it is the 67\% SD phase's nPDF that has equally good
agreement with experiment [Fig.~\ref{Fig:PDF}(d)], while
the zigzag JT (lowest energy state) phase's nPDF shows considerable deviations
[Fig.~\ref{Fig:PDF}(b)].
In the bond-disproportionated phases [Figs.~\ref{Fig:PDF}(b)-(e)],
the two nPDF peaks around 2~{\AA} are particularly sensitive
to the presence of SD NiO$_6$ sites and their evolution
from (b) to (e) is easy to understand.
Indeed, the relative heights of these peaks reflect the ratio
between the numbers of short and long nearest-neighbor Ni-O bonds,
$N_{\text{short}}/N_{\text{long}}$.
As the fraction of SD sites grows,
$N_{\text{short}}/N_{\text{long}}$
gradually decreases from 2 in the fully JT phases to 1
in the fully SD phase (Table~\ref{T:comp})
and so does the intensity of the first
nPDF peak relative to that of the second.
Furthermore, only by having SD sites in the system
can one accurately reproduce the measured position of the second peak
or, equivalently, the average length
of the long Ni-O bonds (Table~\ref{T:comp}).
We would like to emphasize that these results
have been obtained with keeping the nPDF simulations
as {\abinitio} as possible,
which is a step forward compared with the original analysis
of Chung~{\it et al.} in Ref.~\onlinecite{Chung05}.
We also note that, in general, the
fraction of SD octahedra may probably vary from one sample to another
depending on their thermal histories and growth conditions,
which should be a subject of further investigations.

As for the powder diffraction experiments, we
find that the 67\% SD phase gives a diffraction
pattern that is strikingly similar to that
of the rhombohedral phase (Fig.~\ref{Fig:powder}).
Moreover, our general observation is
that all bond-disproportionated phases
with a preserved $C_3$ symmetry of the lattice
tend to have very similar diffraction patterns\cite{powder}.
In other words, the powder diffraction pattern of {\linio}
appears to be barely sensitive to Ni-O bond disproportionation
as long as the $C_3$ symmetry is preserved,
which in fact can explain the long-standing confusion around
the {\linio} crystal structure.

\section{Discussion}

\subsection{Role of entropy}

\begin{figure*}
\centering
\includegraphics[width=0.98\textwidth]{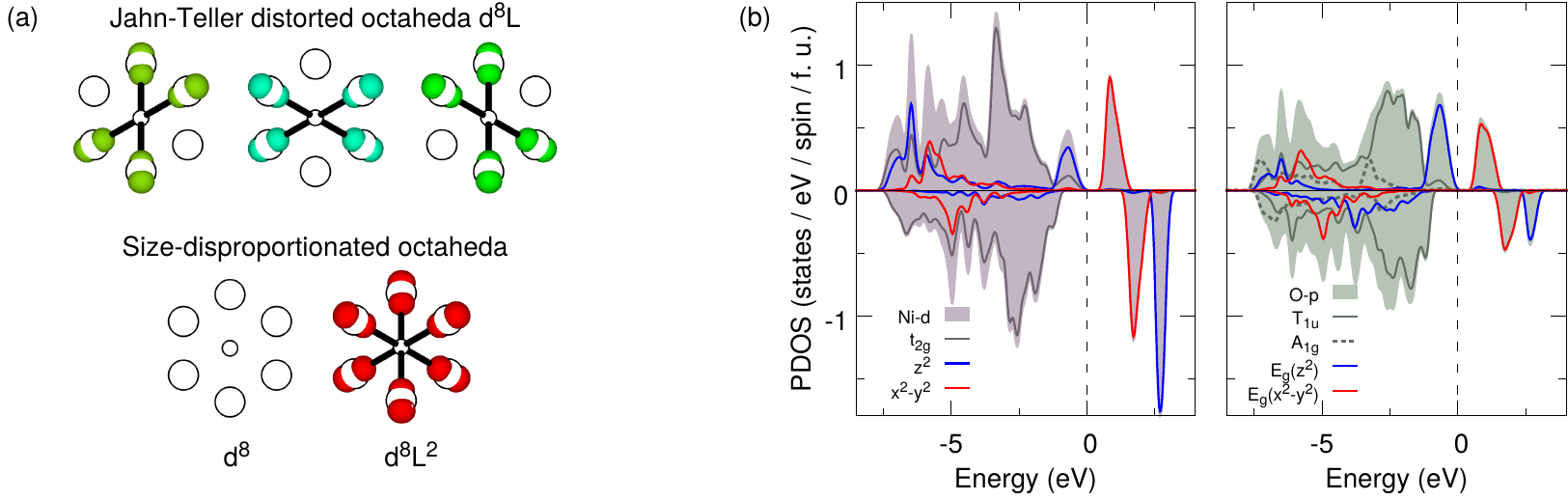}
\caption{(a) The five building block configurations of
the NiO$_6$ octahedra in {\linio}: the three
JT distorted configurations and the two SD configurations.
In color shown are the density iso-surfaces
of holes occupying $E_g$ oxygen molecular orbitals.
(b) Orbitally projected densities of states (PDOS) 
in the zigzag JT phase of {\linio}.
The right panel features projections onto
oxygen molecular orbitals of various symmetries.
Both plots are normalized to a single NiO$_6$ octahedron
with only O-$p_{\sigma}$ orbitals participating.}
\label{Fig:pdos}
\end{figure*}

At this point, the question arises
as to why {\linio} would adopt
a metastable phase like the 67\% SD one
rather than the lowest energy JT zigzag phase,
which is what the above comparison with experiment strongly suggests,
and this is where the entropy factor comes into play.
In fact, given the accuracy limitations of the LDA+U method,
the only robust conclusion that comes from our total
energy calculations
is that there exist phases with varying amounts of
JT distorted and SD NiO$_6$ octahedra whose energies
lie within only 20 to 30~meV/f.~u. above the ground-state energy.
While the ferro JT, 67\% SD, 50\% SD, and 100\% SD phases are just
some examples of such phases, further examples
can be easily come up with and tested in a calculation
by exploring various ways of positioning
the five {\it building block} configurations of a NiO$_6$ octahedron
on a triangular lattice:
the three JT distorted 
and the two SD configurations [Fig.~\ref{Fig:pdos}(a)].
With so many metastable phases being so close in energy,
a state where the system is able to fluctuate between
them can have a high enough entropy to be stabilized
at ambient temperatures by the entropy term in the free energy.
In this fluctuating state, each NiO$_6$ site can assume
any of the building block configurations, and in an extreme
limit would do it in a random way.
In reality, of course, there will be constraints
due to local correlations between different types of sites
mostly determined by Pauling’s principle of electroneutrality\cite{Pauling48}.
Thus, an expanded NiO$_6$ octahedral site with electronic configuration
$d^8$ would ``attract'' collapsed $d^8\underline{L}^2$ sites
(each having $2e$ less charge due to the 2 ligand holes than the reference $d^8$ site)
in its immediate vicinity, but due to the triangular geometry those will be somehow
diluted by $d^8\underline{L}$ and/or $d^8$ sites.
One can envision this mechanism driving a propagation
of local $d^8$, $d^8\underline{L}$, and $d^8\underline{L}^2$
configurations from a nucleation center
ensured by the entropy to be as random as possible at a given temperature.
We believe that the final entropy-stabilized state
will involve local distributions of $d^8$, $d^8\underline{L}$,
and $d^8\underline{L}^2$ sites
corresponding to all and any of those depicted in Fig.~\ref{Fig:str}.
We can in fact make the discussion more quantitative
by computing and comparing the configurational entropy of
this entropy-stabilized state and that of the
lowest-energy zigzag Jahn-Teller phase.
As detailed in the Appendix,
the impact of the vibrational (or phonon) free energy $F^{\text{phon}}$
on the stability of various {\linio} phases
is small compared to that of the configurational entropy
and therefore will not be considered here.
The configurational entropy of the zigzag Jahn-Teller phase
is due to spin fluctuations between the $S_z=\pm \frac12$
states of the Ni ions' spins:
$S^{\text{conf}} = Nk_{\text B} \ln 2
= 0.693 Nk_{\text B}$,
where $N$ is the number of formula units
in the crystal. For the entropy-stabilized state,
let us assume for simplicity that the $d^8$, $d^8\underline{L}$,
and $d^8\underline{L}^2$ sites
have equal concentrations of $1/3$, and that they
are randomly distributed.
In addition, the JT $d^8\underline{L}$ sites have degeneracy of $g_1=6$ corresponding
to three possible orientations of the long Ni-O bond [see Fig.~\ref{Fig:pdos}(a)]
and two $S_z$ components of spin $S=\frac12$, while the large octahedron
sites $d^8$ have degeneracy of $g_2=3$ corresponding to three
$S_z$ components of spin $S=1$. Neglecting local correlations
between the different types of sites,
which otherwise would somewhat reduce our estimate of
$S^{\text{conf}}$, we get that
$S^{\text{conf}} =
k_{\text B} \ln(3^N (g_1^{N/3} g_2^{N/3})) = 2.062 Nk_{\text B}$.
This upper bound is in fact very large and exceeds the value
of $\ln5=1.61$ in five-component mixtures at equal concentrations,
which are conventionally considered ``high-entropy'' materials\cite{Murty2014}.
We note that, aside from the chemical
composition aspects used in the design of high-entropy
alloys\cite{Rost15,Sarkar18},
our analysis puts a new twist on the determination of the entropy
to include Jahn-Teller as well as electronic and local disproportionation.

It would be very interesting to
look for experimental signatures of the
glass transition in {\linio}, at which the electronic
and orbital fluctuations discussed above should freeze. For example, it was found
that the heat capacity of {\linio} is anomalously enhanced below 300~K\cite{Kawaji02}.
Clearly, further heat capacity measurements exploring much higher
temperatures are strongly desired in order to observe the full glass transition.
Another experimental observation which is strongly
supportive of the glassy state in {\linio}
is the spin-glass like transition observed at around 20~K.
Indeed,
having the Ni sites electronically in a random mixture 
of $d^7\underline{L}$, $d^8\underline{L}^2$,
and $d^8$ configurations
means having a random mixture of $S=\frac12$, $S=0$, and $S=1$ spin magnetic moments,
respectively, with similarly randomized exchange interactions between them,
which is an important prerequisite for the formation of
a spin glass.

From the application point of view,
the high configurational entropy of the
glassy state in {\linio} is very beneficial in terms of battery operation
as the system has many ways to adjust itself locally
to a missing Li ion, which is apparently what makes {\linio} such a
good cathode. In this regard, further increase of entropy upon
chemical substitution of Ni with Co and Mn
may be
the main reason for the increased
stability of LiNi$_{1-x-y}$Co$_x$Mn$_y$O$_2$ cathodes.
Another example of this concept can be found in a recent study
on multi-component monoxides which demonstrates that five-component
systems perform much better in batteries compared with four- or less component
systems and attributes this effect to entropy stabilization\cite{Sarkar18}.

\subsection{Oxygen holes}

Finally, in order to better understand the microscopic nature of
the disordered glassy state in {\linio},
let us examine its
electronic structure.
We first note that LDA+U
correctly describes {\linio} as a negative
charge-transfer system\cite{Kuiper89},
as far as the Ni oxidation state is concerned.
Figure~\ref{Fig:pdos}(b) compares the
Ni-$3d$ and O-$2p$ orbitally projected densities of states (DOS)
in the zigzag JT phase. One can
see a considerable amount of the O-$2p$ character
in the states above the Fermi level, indicating
that the true Ni oxidation state is far less than 3+.
We find close to 8 electrons in the Ni-$3d$ shell
inside the muffin-tin sphere.
What obscures the picture, though, is the strong hybridization
between the Ni-$e_g$ orbitals and the oxygen-$2p$ orbitals of a respective
symmetry. As one can see
in the right panel of Fig.~\ref{Fig:pdos}(b), it results in holes occupying
$E_g$-symmetric molecular-like orbitals formed by the oxygen-$2p_{\sigma}$ orbitals
in an octahedral cage.
Although LDA+U is a mean-field method not capable of properly describing
a many-body wave-function,
we still can associate the states observed in our calculations
on the expanded, JT distorted,
and collapsed octahedra
with, respectively, the $t_{2g}^6e_g^2$, $(t_{2g}^6e_g^2)\underline{L}$,
and $(t_{2g}^6e_g^2)\underline{L}^2$ configurations.
Hole density iso-surfaces for each configuration
are schematically shown in Fig.~\ref{Fig:pdos}(a).
Note that a single hole occupying the $x^2-y^2$-symmetric
$E_g$ molecular orbital results in a {\it shortening}
of only four Ni-O bonds, while two holes occupying
both the $x^2-y^2$- and the $z^2$-symmetric $E_g$ orbitals result in
a {\it collapse} of all Ni-O bonds.
We further find 1.68, 0.90, and 0.08~$\mu_{\text{B}}$ for the magnetic
moments of the Ni ions inside the expanded, JT distorted, and collapsed
octahedra, respectively.
Again, this is a result of LDA+U being a mean-field method;
in a true many-body calculation, one would find
for $(t_{2g}^6e_g^2)\underline{L}^2$, for example,
that the $S=1$ moment on the nickel forms a singlet state with the $S=1$
moment on the oxygen molecular orbital and the total moment
on the octahedron is zero.

\begin{figure}
\centering
\colorbox{white}{
\includegraphics[width=0.5\textwidth]{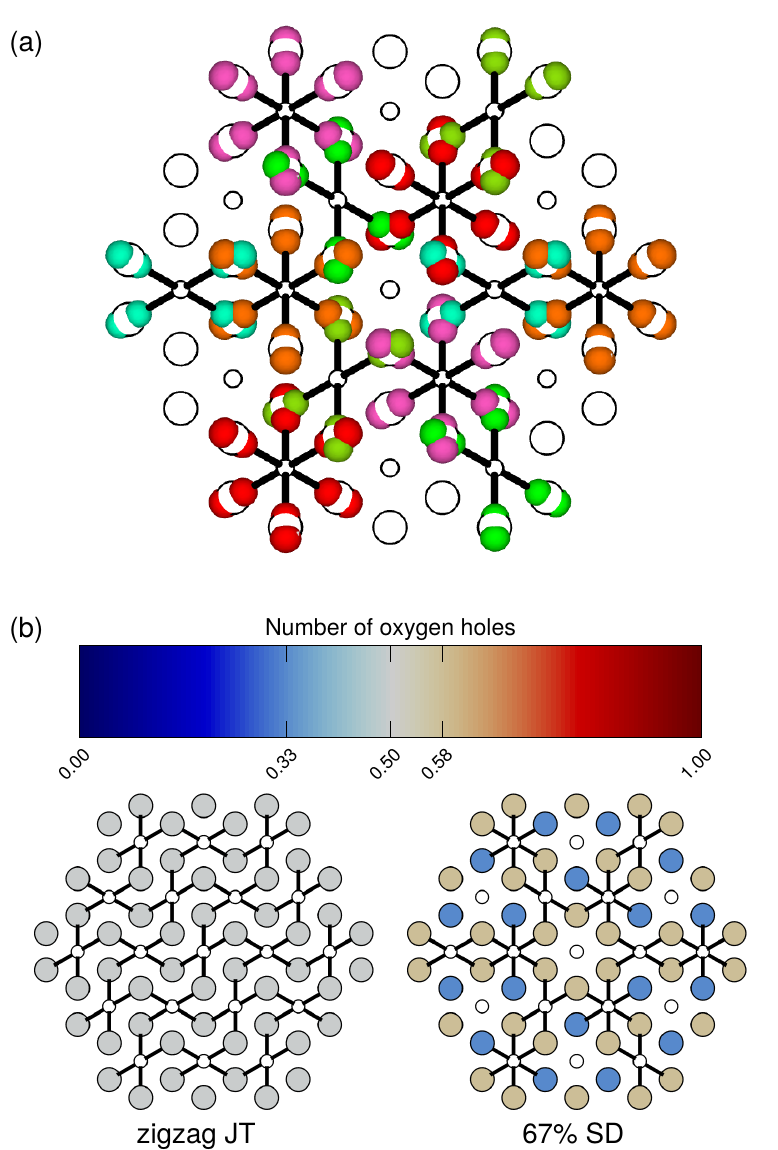}
}
\caption{Oxygen charge disproportionation in {\linio}.
(a) A schematic illustration of the density iso-surfaces of holes occupying
individual oxygen molecular orbitals in the 67\% SD phase.
(b) Oxygen charge disproportionation
in the 67\% SD phase versus a homogeneous hole distribution
in the zigzag JT phase.}
\label{Fig:MO}
\end{figure}

Yet another peculiar property of the SD phases
in {\linio} is that there occur both bond disproportionation and
{\it oxygen charge disproportionation}.
This is in contrast with the rare-earth nickel perovskites,
where only {\it bond disproportionation} occurs,
which is a result of a
different lattice geometry and a different
average number of holes per oxygen octahedron (three in {\linio}
versus two in the rare-earth nickel perovskites).
In Fig.~\ref{Fig:MO}(a), we schematically
show density iso-surfaces of holes occupying individual molecular orbitals
in the 67\% SD phase of {\linio}.
One can clearly distinguish oxygen sites where neighboring molecular orbitals
overlap from those where they do not. As a result, the two types
of oxygen sites have different
hole concentrations, which is further illustrated in Fig.~\ref{Fig:MO}(b).

The extended nature of the oxygen molecular orbitals
and the fact that they can overlap result
in a strong nearest-neighbor interaction between holes,
which is generally known to facilitate charge ordering effects.
Although formation of a glassy state
may in general depend on various factors, we can speculate that 
in such an environment
localized charge impurities can serve as
nucleation centers for charge disproportionation.
In {\linio}, charge impurities are readily available
in a form of doped electrons entering the Ni layers
as a result of the unavoidable deviation from stoichiometry\cite{Arai95}.
The doped electrons would attract holes according to Pauling's principle\cite{Pauling48}
and thus locally modify the energetics in favor of charge- (and size-)
disproportionated phases. Further investigations are, however,
required in order to validate this scenario.

We believe that spectroscopic probes, such as x-ray
absorption spectroscopy (XAS), could be crucial in verifying the mixed size-disproportionated
state in {\linio}. Especially interesting would be to compare spectroscopic
data from {\linio} and its sister compound {\nanio} where the ground state is
contrastingly uniform. Unfortunately,
unambiguous interpretation of the existing
Ni $2p$ XAS data\cite{Abbate91,Veenendaal94,Montoro99} is not easy
due to questionable sample quality and also the fact that
their proper theoretical description requires to go beyond
the standard single cluster calculations\cite{Green16}.
We hope therefore that our theoretical proposal of
the glass-like electronic ground state of {\linio}
will stimulate experimental efforts in preparing and systematically
measuring high-quality {\linio} and {\nanio} samples complemented with
a proper theoretical analysis of spectroscopic
data in the spirit of Ref.~\onlinecite{Green16}.
We also note that in the case of the negative
charge-transfer gap rare-earth nickelates the detailed analysis of
their XAS data identifying the origin
of the two-peak structure was only made after the single
crystalline epitaxial thin film studies with resonant inelastic x-ray
scattering by Bisogni {\etal}\cite{Bisogni16}.
This kind of study on epitaxial thin films of {\linio} and {\nanio} would
be extremely helpful.

\section{Conclusion}
In summary, we have used electronic structure methods
to demonstrate that the
Li-ion battery material {\linio}
is in a high-entropy charge-glass like state
characterized by a disordered mixture of JT distorted
and SD NiO$_6$ octaherda.
This state is associated with charge disproportionation
on the oxygen rather than on the nickel
atoms due to {\linio} being in the negative charge-transfer regime.
Supported by rigorous calculations,
our proposal explains
the previously highly debated
nPDF and powder diffraction measurements on {\linio}
as well as its enhanced heat capacity and the 
spin-glass like behavior.
Most importantly, however,
we can conclude that it is the glassy nature that
renders {\linio} its extraordinarily high reversible
capacity.
We believe that recognizing the role of entropy
in stabilizing cathodes during the charge and discharge
cycles can greatly advance battery research.

\section*{Appendix}
\subsection{Dependence of energy on a spin configuration}
\begin{figure}
\centering
\includegraphics[width=0.45\textwidth]{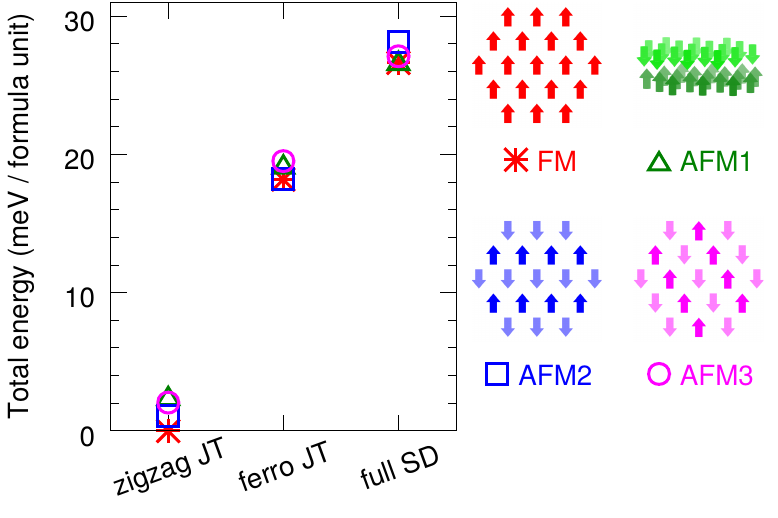}
\caption{Relative total energies
as a function of a structural phase
and Ni spin configuration in {\linio}.
FM stands for the ferromagnetic spin configuration.
The anti-ferromagnetic configuration AFM1 corresponds
to alternating layers of up and down spins.
The anti-ferromagnetic configurations AFM2 and AFM3
correspond to two different arrangements of
up and down spins within a layer.
Note that AFM2 and AFM3 are not equivalent
due to the broken $C_3$ symmetry in the considered structural phases.}
\label{Fig:s1}
\end{figure}

Figure~\ref{Fig:s1} demonstrates that the energy
differences between different
structural phases of {\linio}
are very weakly dependent on the Ni spin configuration.
In this plot, we have considered
the zigzag JT, ferro-orbital JT, and fully
size-disproportionated structural phases
as well as one ferromagnetic and three anti-ferromagnetic
Ni spin configurations labeled as FM, AFM1, AFM2, and AFM3.

\subsection{Calculation of thermodynamic properties}
\begin{figure*}
\centering
\includegraphics[width=0.9\textwidth]{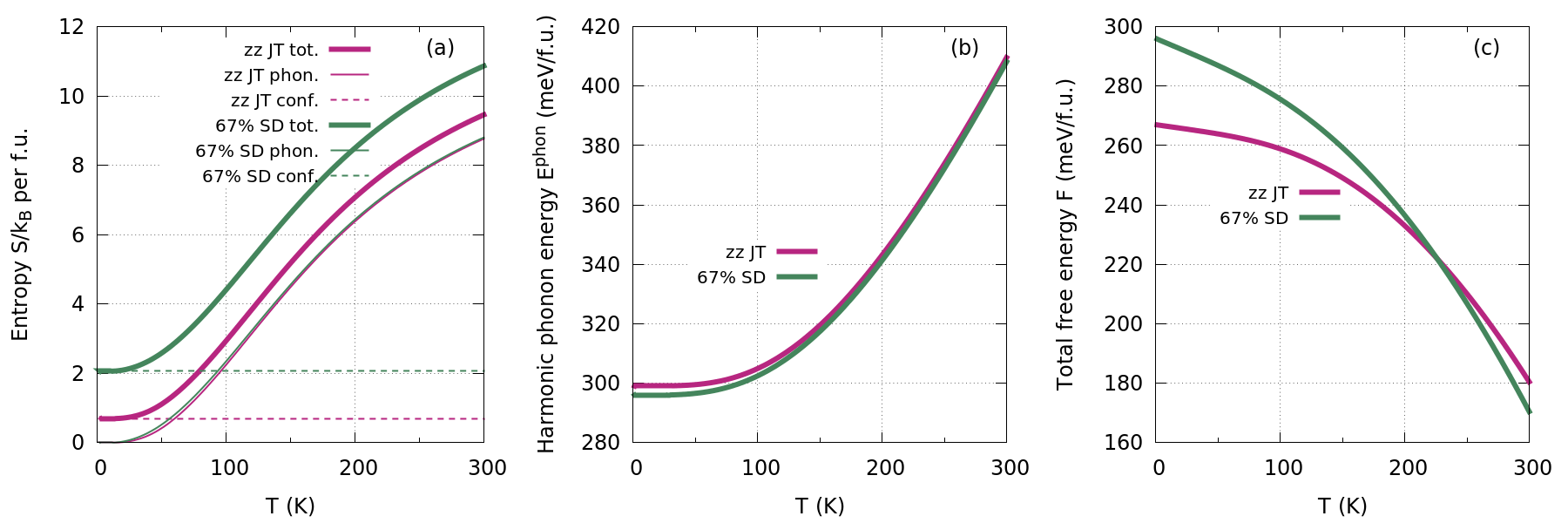}
\caption{
Temperature dependence of the
thermodynamic properties of {\linio} in the 
zigzag Jahn-Teller (zz JT) and the 67\% size-disproportiotaled (67\% SD) phases:
(a) configurational, vibrational, and total entropies $S$;
(b) harmonic phonon energy $E^{\text{phon}}$;
(c) total free energy $F$.
}
\label{Fig:s2}
\end{figure*}

In order to estimate how important is the impact of
the vibrational free energy on the stability
of various {\linio} phases, we have computed the vibrational free energies
$F^{\text{phon}} = E^{\text{phon}} - TS^{\text{phon}}$
of the zigzag JT and the 67\% SD phases of {\linio}
using the phonon spectra obtained with VASP and Phonopy via
the dynamical matrix method. The computed
vibrational entropies $S^{\text{phon}}$, the harmonic phonon energies
$E^{\text{phon}}$,
as well as the total entropies and total free energies
$F = E^{T=0} + F^{\text{phon}} - TS^{\text{conf}}$
are shown in Fig.~\ref{Fig:s2} as a function of temperature $T$.
$E^{T=0}$ is the $T=0$ energy, which is
0 and +32.1 meV/f.u. in the zigzag JT and the 67\% SD phases,
respectively. It is assumed that the configurational
entropy of the 67\% SD phase is that of the entropy-stabilized state
with equal concentrations of $d^8$, $d^8\underline{L}$,
and $d^8\underline{L}^2$ sites in order to illustrate
its significant contribution compared with the phonon contribution.

%
%
%
%

\section*{Acknowledgments}
The authors are grateful to Daniel Khomskii for
insightful discussions.
K.~F. thanks Jae-Ho~Chung and Thomas~E.~Proffen for sharing their
neutron data and Leopoldo Suescun for helping
with GSAS simulations.

\bibliography{ref}

\begin{thebibliography}{43}
\bibitem[{pow()}]{powder}
\bibinfo{note}{For instance, such is a JT phase where the direction of the long
  Ni-O bond in an octahedron alternates between successive layers.}
\end{thebibliography}

\end{document}